\documentclass{amsproc}
\usepackage{amsfonts}

\theoremstyle{plain}

\numberwithin{equation}{section}
\input{tcilatex}

\begin{document}
\title[]{THE LIGHT FRONT GAUGE PROPAGATOR: THE STATUS QUO}
\author{A. T. Suzuki$^{a}$}
\address{$^{a,b}$Instituto de F\'{\i}sica Te\'{o}rica-UNESP, 01405-900\\
S\~{a}o Paulo, Brazil.}
\author{J.H.O. Sales$^{b}$}

\begin{abstract}
At the classical level, the inverse differential operator for the quadratic
term in the gauge field Lagrangian density fixed in the light front through
the multiplier $(n\cdot A)^{2}$ yields the standard two term propagator with
single unphysical pole of the type $(k\cdot n)^{-1}$. Upon canonical
quantization on the light-front, there emerges a third term of the form $%
(k^{2}n^{\mu }n^{\nu })(k\cdot n)^{-2}$. This third term in the propagator
has traditionally been dropped on the grounds that is exactly cancelled by
the \textquotedblleft instantaneous\textquotedblright\ term in the
interaction Hamiltonian in the light-front. Our aim in this work is not to
discuss which of the propagators is the correct one, but rather to present
at the classical level, the gauge fixing conditions that can lead to the
three-term propagator. It is revealed that this can only be acomplished via
two coupled gauge fixing conditions, namely $n\cdot A=0=\partial \cdot A$.
This means that the propagator thus obtained is doubly transversal.
\end{abstract}

\date{}
\maketitle

\section{INTRODUCTION}

As early as 1970 with J.B.Kogut and D.E.Soper [1] and a little later with
E.Tomboulis in 1973 [2], light-front gauge propagator for Abelian and
Non-Abelian gauge fields derived via canonical quantization was known to
have a (third) term proportional to $(k^2 n_\mu n_\nu)(k\cdot n)^{-2}$.
According to the latter, \emph{``The third term represents an instantaneous
``Coulomb''-type interaction.''} Moreover, he (see also [3]) showed then
that \emph{``We will now show that all graphs representing the Coulomb term
... precisely cancel the contributions from the last term of the propagator
... so that we are left with an effective interaction Hamiltonian density
... i.e, the two usual vertices, and a propagator...''} where the so-called 
\emph{effective} propagator is the traditional two-term light-front
propagator: 
\begin{equation}  \label{reduced}
G^{\mu\nu\mathrm{ab}}(k)=-\frac{\delta^{\mathrm{ab}}}{k^2+i\varepsilon}\left[%
g^{\mu\nu}-\frac{k^\mu n^\nu+k^\nu n^\mu}{k\cdot n}\right]
\end{equation}
More recently, P.P.Srivastava and S.J.Brodsky [4] rederived the three-term
``doubly transverse gauge propagator'' 
\begin{equation}  \label{full}
G^{\mu\nu\mathrm{ab}}(k)=-i\frac{\delta^{\mathrm{ab}}}{k^2+i\varepsilon}%
\left[g^{\mu\nu}-\frac{k^\mu n^\nu+k^\nu n^\mu}{k\cdot n}+\frac{k^2 n^\mu
n^\nu}{(k\cdot n)^2}\right],
\end{equation}
for QCD in the framework of Dyson-Wick S-matrix expansion with a \emph{BRS
symmetric} Lagrangian density. They then apply the Dirac method of
implementing the field constraints (first and second classes) to obtain the
interaction Hamiltonian from which the canonical quantization is performed
via correspondence principle between Poisson brackets and Dirac commutators
for the field operators. Their derivation clearly shows the conspicuous
instantaneous interaction terms (the so-called tree-level \emph{seagull}
diagram terms) present in the interaction Hamiltonian in the light-front.
Their explicit calculations for the electron-muon scattering in the Abelian
QED theory in the light-front as well as the one-loop $\beta$-function for
the non-Abelian Yang-Mills fields, with gluon vacuum polarization tensor,
three-point vertex functions and gluon self-energy corrections from the
quark loop, show us the subtle cancellations that come to play a crucial
role into the game of light-front renormalization program with instantaneous
interaction terms in the Hamiltonian and the third term of the gluon
propagator.

On the other hand, if one uses the classical approach of inverting the
differential operator sandwiched between the quadratic term in the
Lagrangian density plus the gauge fixing term of the form $(n\cdot A)^2$ in
order to obtain the gauge field propagator, the result is straightforwardly
given by (\ref{reduced}). There is no way - classically - to arrive at (\ref%
{full}) with only the gauge fixing Lagrangian of the form $(n\cdot A)^2$.
This means that, as it stands, there is an anomaly between the classical and
the quantum propagator.

\section{CLASSICALLY DEDUCIBLE THREE-TERM L.F. PROPAGATOR}

Since at the classical level we just look for the inverse operator
sandwiched between the quadratic term in the Lagrangian density plus the
gauge fixing term, in order to get a three-term propagator we need to
incorporate not only the usual $n\cdot A=0$ condition into the gauge fixing
part, but couple it to the Lorentz condition $\partial \cdot A=0$. The
reason why we need the latter condition becomes clear when one understands
that the Lorentzian condition coupled to the former gauge condition is
nothing more than the constraint equation for the unphysical field component 
$A^{-}$, which is not a dynamical variable in the light-front formalism.
Note that this does not remove too many degrees of freedom from the gauge
fields as one would naively think, but that both of those two are in fact
necessary to completely fix the gauge in the light-front with no residual
gauge freedom left. In a recent work, we have shown how this can be
accomplished [5] via considering one Lagrange multiplier of the form $%
(n\cdot A)(\partial \cdot A)/\alpha$, where $\alpha$ is the single gauge
fixing parameter.

In this work we show that the one gauge fixing term above referred to can be
generalized to a two term general gauge fixing term of the form $(n\cdot
A)^2/\alpha+(\partial \cdot A)^2/\beta$, where now $\alpha$ and $\beta$ are
two independent gauge fixing parameters, yielding the same result, namely,
the three-term, ``doubly transverse propagator'' (\ref{full}). .

The Lagrangian density for the vector gauge field (for simplicity we
consider an Abelian case) is given by 
\begin{equation}
\mathcal{L}=-\frac{1}{4}F_{\mu \nu }F^{\mu \nu }-\frac{1}{2\beta }\left(
\partial _{\mu }A^{\mu }\right) ^{2}-\frac{1}{2\alpha }\left( n_{\mu }A^{\mu
}\right) ^{2}=\mathcal{L}_{\mathrm{E}}+\mathcal{L}_{GF}  \label{2}
\end{equation}

By partial integration and considering that terms which bear a total
derivative don't contribute and that surface terms vanish since $
\lim\limits_{x\rightarrow \infty }A^{\mu }(x)=0$, we have 
\begin{equation}
\mathcal{L}_{\mathrm{E}}=\frac{1}{2}A^{\mu }\left(\partial^2 g_{\mu \nu
}-\partial _{\mu }\partial _{\nu }\right) A^{\nu }  \label{3}
\end{equation}
and 
\begin{eqnarray}
\mathcal{L}_{GF}&=&-\frac{1}{2\beta }\partial _{\mu }A^{\mu }\partial _{\nu
}A^{\nu }-\frac{1}{2\alpha }n_{\mu }A^{\mu }n_{\nu }A^{\nu } \\
&=&\frac{1}{2\beta}A^{\mu }\partial _{\mu }\partial _{\nu }A^{\nu }-\frac{1}{
2\alpha }A^{\mu }n_{\mu }n_{\nu }A^{\nu }  \label{4}
\end{eqnarray}
so that 
\begin{equation}
\mathcal{L}=\frac{1}{2}A^{\mu }\left( \partial^2 g_{\mu \nu}-\partial_{\mu
}\partial _{\nu }+\frac{1}{\beta }\partial _{\mu }\partial _{\nu }-\frac{1}{
\alpha }n_{\mu }n_{\nu }\right) A^{\nu }  \label{5}
\end{equation}

To find the gauge field propagator we need to find the inverse of the
operator between parenthesis in (\ref{5}). That differential operator in
momentum space is given by: 
\begin{equation}
O_{\mu \nu }=-k^{2}g_{\mu \nu }+k_{\mu }k_{\nu }-\theta k_{\mu }k_{\nu
}-\lambda n_{\mu }n_{\nu }\,,  \label{6}
\end{equation}
where $\theta =\beta ^{-1}$ and $\lambda =\alpha ^{-1}$, so that the
propagator of the field, which we call $G^{\mu \nu }(k)$, must satisfy the
following equation: 
\begin{equation}
O_{\mu \nu }G^{\nu \lambda }\left( k\right) =\delta _{\mu }^{\lambda }
\label{7}
\end{equation}

$G^{\nu \lambda }(k)$ can now be constructed from the most general tensor
structure that can be defined, i.e., all the possible linear combinations of
the tensor elements that composes it [6]: 
\begin{eqnarray}
G^{\mu \nu }(k) &=&g^{\mu \nu }A+k^{\mu }k^{\nu }B+k^{\mu }n^{\nu }C+n^{\mu
}k^{\nu }D+k^{\mu }m^{\nu }E+ \\
&&+m^{\mu }k^{\nu }F+n^{\mu }n^{\nu }G+m^{\mu }m^{\nu }H+n^{\mu
}m^{\nu}I+m^{\mu }n^{\nu }J  \label{8}
\end{eqnarray}
where $m^\mu$ is the light-like vector dual to the $n^\mu$, and $A$, $B$, $C$
, $D$, $E$, $F$, $G$, $H$, $I$ and $J$ are coefficients that must be
determined in such a way as to satisfy (\ref{7}). Of course, it is
immediately clear that since (\ref{5}) does not contain any external
light-like vector $m_{\mu }$, the coefficients $E=F=H=I=J=0$ straightaway.
Then, we have 
\begin{eqnarray}
A&=&-(k^{2})^{-1}  \label{9} \\
(k\cdot n)(1-\theta )G-\theta k^{2}D&=& 0  \label{10} \\
(-k-\lambda n^{2})G-\lambda (k\cdot n)D-\lambda A&=& 0  \label{11} \\
-(k^{2}+\lambda n^{2})C-\lambda (k\cdot n)B&=&0  \label{12} \\
(1-\theta )A-\theta k^{2}B+(1-\theta )(k\cdot n)C&=&0  \label{13}
\end{eqnarray}

From (\ref{10}) we have 
\begin{equation}
G=\frac{k^{2}}{(k\cdot n)(\beta -1)}D  \label{14a}
\end{equation}
which inserted into (\ref{11}) yields 
\begin{equation}
D=\frac{-(k\cdot n)(\beta -1)}{(\alpha k^{2}+n^{2})k^{2}+(k\cdot
n)^{2}(\beta -1)}A  \label{15}
\end{equation}

From (\ref{12}) and (\ref{13}) we obtain 
\begin{equation}
B=\frac{-(\alpha k^{2}+n^{2})}{k\cdot n}C  \label{16}
\end{equation}
and 
\begin{equation}
C=\frac{-(\beta -1)(k\cdot n)}{(\alpha k^{2}+n^{2})k^{2}+(k\cdot
n)^{2}(\beta -1)}A=D
\end{equation}

In the light-front, $n^{2}=0$ and taking the limits $\alpha,\,\beta
\rightarrow 0$, we have 
\begin{eqnarray}
A&=&\frac{-1}{k^{2}} \\
B&=&0 \\
C=D&=&\frac{1}{k^{2}(k\cdot n)} \\
G&=&\frac{-1}{(k\cdot n)^{2}}
\end{eqnarray}

Therefore, the relevant propagator in the light-front gauge is: 
\begin{equation}
G^{\mu \nu }(k)=-\frac{1}{k^{2}}\left\{ g^{\mu \nu }-\frac{k^{\mu }n^{\nu
}+n^{\mu }k^{\nu }}{k\cdot n}+\frac{n^{\mu }n^{\nu }}{(k\cdot n)^{2}}
k^{2}\right\} \,,  \label{17}
\end{equation}
which has the outstanding third term commonly referred to as \emph{contact
term}. This procedure eliminates the problem of the existing anomaly between
the classical and quantum derivations for the light-front gauge propagator.

\section{Conclusions}

We have shown that at the classical level we can introduce two Lagrange
multipliers in the Lagrangian density that is consistent with gauge fixing
in the light-front. No excess degrees of freedom is eliminated with this
formalism since the coupled conditions are such that the Lorentzian
condition yields nothing more than a constraint equation for the
non-dynamical variable $A^{-}$. This means that with both these coupled
conditions, the ensuing gauge field is entirely defined in its dynamical
transverse degrees of freedom, i.e., there is no residual gauge freedom
left. Moreover, the consistency of the procedure is enhanced by the fact
that the propagator thus deduced is the same as the one obtained via
canonical quantization in the light-front (no anomaly). The reason why this
is so can be seen from the fact that since $n\cdot A=0=\partial \cdot A$ it
follows that $(n\cdot A+\partial \cdot A)^2=0$ from which $(n\cdot
A)^2+(\partial \cdot A)^2=-2(n\cdot A)(\partial \cdot A)$ [7]. Of course, no
attempt is here made to discuss whether one should work in the usual
two-term reduced (or sometimes called effective) propagator or in the
three-term (which we call full) propagator. Our aim here was solely to solve
the anomaly problem and pinpoint a solution to the classical problem when
confronted with the quantum derivation.

\vskip1cm \noindent

\section*{ACKNOWLEDGEMENTS}

One of the authors (ATS) wishes to thank the ILCAC and S.Dalley as the chair
of the Local Organizing Committee for the organization of the meeting, and
Prof. W.J.Stirling for the hospitality and partial sponsoring while at IPPP.
ATS thanks partial support from CNPq (Bras\'{\i}lia, DF) and FUNDUNESP
(S\~ao Paulo, SP). JHOS thanks FAPESP (S\~ao Paulo, SP) for financial
support.

\end{document}